\documentclass[prl,aps,twocolumn,showpacs]{revtex4}
\usepackage{graphicx}

\newcommand{\half}{\textstyle{1\over 2}}
\newcommand{\be}{\begin{equation}}
\newcommand{\ee}{\end{equation}}
\newcommand{\beqa}{\begin{eqnarray}}
\newcommand{\eeqa}{\end{eqnarray}}
\newcommand{\nonu}{\nonumber}
\newcommand{\re}{{\rm e}}
\newcommand{\pr}{\prime}
 \newcommand{\lan}{\langle}
  \newcommand{\ran}{\rangle}
  \begin{document}
  
  \title{Dissipation and decoherence in a quantum oscillator}

   \author{Vinay Ambegaokar}
\affiliation{Laboratory of Atomic and Solid State Physics, Cornell 
University, Ithaca, New York 14853}
\date{August 9, 2005}

 \begin{abstract}
 The time development of the reduced density matrix for a quantum oscillator damped by coupling it to an ohmic environment is calculated via an identity of the  Debye-Waller form.  Results obtained some years ago by Hakim and the author in the free-particle limit  \cite{VHVA} are thus recovered.  The evolution of a free particle in a prepared initial state is examined, and a previously published exchange  \cite{FOC,GvDA} is illuminated with figures  showing no decoherence without dissipation.
  \end{abstract}
   \maketitle

 \section{ I. Introduction} 
 In a recent paper in this journal, van Kampen\cite{NGK} has re-examined dissipation and noise in a quantum oscillator, treating it as a sub-system coupled to an environment.  In working  out his model, he has introduced some nice methods.  They are slightly simplified and modified in this paper to revisit the closely related problem of quantum coherence and decoherence. 
 
It is worth emphasizing that the harmonic oscillator is particularly simple, so that the analysis given here is not generalizable to more interesting systems.  However, it may make up in explicitness what it lacks in generality.  Indeed, the ultimate aim of this research is to further clarify the ``positivity problem" in time dependent quantum statistical equations.\cite{VA, SP}   The present work is a first step in that direction, in the hope that deriving largely known results in a simple way will clear the path. 
 
This simplicity may offer amusement if not instruction  to my long-time friends and  colleagues Jim Langer and Pierre Hohenberg,  despite the burdens of their high offices.   It is a  pleasure and honor to dedicate the paper to them.

 \section{II. Model and Preliminaries}
 The model is described by the Hamiltonian 
 \be\label{1d}
 H =\half  [P_0^2 +\Omega_0^2 Q_0^2 ]+ \half \sum_k \{P_k^2 + \omega_k^2[Q_k + (\alpha_k Q_0/\omega_k^2)]^2\} ,
 \ee
 where the oscillator labeled $0$ will be called ``the sub-system"  and the others ``the environment."
  The quantum mechanical (and classical) equations of  motion obtained from Eq. (1) are
 \beqa
  \ddot Q_0 +\big [\Omega_0^2 + \sum_k (\alpha_k / \omega_k)^2\big]Q_0& +& \sum_k \alpha_k Q_k= 0\nonu \\
   \ddot Q_k +\omega_k^2 Q_k + \alpha_k Q_0&= &0.
  \eeqa
  Fourier transforming Eq.(2) and eliminating $Q_k$, one  obtains 
  \be
 g^{-1} (\omega + i 0^+) Q_0 = 0, ~~~{\rm where}
 \ee
 \be 
 g^{-1}(z) = z^2 - \Omega_0^2 - \sum_k \alpha_k^2 \big ( {1\over z^2 - \omega_k^2} + {1\over \omega_k^2} \big),
  \ee 
$z$ is a complex variable, and the $i0^+$ in Eq.(3) introduces the causal boundary condition.   As pointed out in ref.\cite{NGK}, where it is called $G$, $g^{-1}(z)$ has zeros on the real $z$ axis corresponding to the normal mode frequencies,  $\omega_\nu$,  of the coupled system of oscillators, and $Q_0(\omega_\nu)$ is the amplitude of the sub-system oscillator in the $\nu$th mode.

Now, ohmic dissipation \cite{LC} requires 
 \be
 {\pi\over 2} \sum_k (\alpha_k^2/ \omega_k)\delta (\omega -\omega_k) \equiv J(\omega)=\eta \omega
 \ee
 for $\omega$ less than an upper cut-off $\omega_c$.   Substituting this form into Eq. (4) yields, for $\omega_c\gg \omega$,
 \beqa
 g^{-1} (\omega + i0^+)&=& \omega^2 - \Omega_0^2 -{2\eta\over \pi} \int_0^{\omega_c} d\bar\omega
 {\omega^2\over (\omega + i0^+)^2 - \bar\omega ^2}\nonu\\
 &=& \omega^2 - \Omega_0^2 + i\omega \eta,
 \eeqa
demonstrating very explicitly that Eqs.  (1) and (5) do indeed construct a linearly dissipative environment, with damping constant $\eta$, without changing the system frequency $\Omega_0$.
 
 In ref.\cite{NGK}, it is noted that the orthogonal normal mode transformation matrix $X$ defined by
 \be
 Q_0= \sum_\nu X_{0\nu} q_\nu ~~~~~Q_k = \sum_\nu X_{k  \nu}  q_\nu,
 \ee   where the $q_\nu$s are the normal  co-ordinates, is obtainable from the Green function given in Eq. (4).  The normalizations
 \beqa
 \sum_\nu q_\nu^2&=&1~~~{\rm and}\nonu\\ Q_0^2 + \sum_k Q_k^2 = Q_0^2 [1 &+&\sum_k \alpha_k^2/(\omega_k^2 - \omega^2)^2]=1,
 \eeqa
show that the amplitudes corresponding to the mode $\nu$, and thus the matrix elements of the transformation, are given by
 \beqa
 {1\over X_{0\nu}} &=& \sqrt{1 + \sum_k\alpha_k^2/(\omega_k^2 - \omega_\nu^2)^2}\nonu\\
 &=&\sqrt{{1\over 2 \omega_\nu } {\partial g^{-1}\over \partial \omega_\nu}}~~~{\rm and}\nonu\\
X_{k \nu} &=& {\alpha_k\over \omega_\nu^2 - \omega_k^2 } X_{0\nu}.
 \eeqa
 One also learns from ref.\cite{NGK} that Eq.(9) may be used to deftly perform sums over normal modes.  The complex function $g(z)$ has poles only on the real axis and thus the spectral representation
 \be
 g(z)=\int_{-\omega_c}^{\omega_c}\frac{d \omega}{2 \pi}\frac{s(\omega)}{z -\omega}.
 \ee
 Using the explicit form for $X_{0 \nu} $ given in the second line of Eq.(9) one obtains a formula that will be useful later in this  work:
 \beqa
 \sum_\nu X_{o\nu}^2 F(\omega_\nu) &=&2\oint {d z\over 2\pi i} z g(z) F(z)\nonu\\  &=&2\int_0^{\omega_c}\frac {d \omega}{2
\pi} \omega s(\omega) F(\omega).
\eeqa
The contour surrounds the real axis, where the function F has been assumed to be regular and zero for $\omega < 0$; it has been evaluated using Eq.(10).

Note also from Eq.(6) and its complex conjugate that for an ohmic environment
 \be
 s(\omega) = {2\omega \eta\over (\omega^2 - \Omega_0^2)^2 + \omega^2 \eta ^2}.
\ee
\section{III. Time evolution of the reduced density matrix}
To carry out the program of this section, initial conditions must be specified.   I shall use those of \cite{FOC} and \cite{GvDA}, which are a special case of ones originated, to the best of my knowledge, in ref.\cite{VHVA}.   Assume that at $t=0$ complete thermal equilibrium  is disturbed by a real  ``aperture function" $\alpha (Q_0)$.  The entire system is then allowed to evolve to time $t$, and projected on to position states of the sub-system.  The resulting reduced density matrix is given by ($\hbar = 1$)
\beqa
&~&~~~~~~~~~~~~~~~~\rho(Q_{0f}^\prime, Q_{0f}^{\prime\prime}, t)\\&&\equiv Tr\{\vert Q_{0f}^{\prime\prime}\rangle\langle Q_{0f}^\prime\vert \re ^{-i H t} \alpha (Q_0) \rho_{th} (H) \alpha (Q_0) \re^{+i H t} \}.\nonu
\eeqa
In this equation, the primed quantities are ordinary numbers,  the unprimed ones operators,   $Tr$ indicates a trace over all states of $H$, and $\rho_{th} \equiv  \exp\{-\beta H \} / Tr\exp\{-\beta H\}$ with $\beta$ the reciprocal temperature.
Fourier transform the aperture function 
\be
\alpha (Q_0) = \int  d a \tilde\alpha (a)\re^{iQ_0 a} ,
\ee
and express the projection operator in Eq.(13) as
\beqa
\vert Q_{0f}^{\prime\prime}\rangle\langle Q_{0f}^\prime\vert&= &\int d u d v f(u,v) \re^{iP_0 u} \re^ {iQ_0 v}\nonu~~~~~ {\rm with}\\ 
f(u, v) &= &{1\over 2\pi} \re^{-i Q_{0f}^\prime v} \delta(Q_{0f}^\pr - Q_{of}^{\pr\pr} -u),
\eeqa
proved by taking matrix elements of both sides, to see that  
\beqa
\rho(Q_{0f}^\prime, Q_{0f}^{\prime\prime}, t)&=&\int dadbdudv \tilde\alpha(b) \tilde\alpha (a) f(u,v)\mathcal T,\nonu\\
~~{\rm with}~~~\mathcal T& \equiv&
\langle\re^{iQ_0b} \re^{iP_0(t)u}\re^{iQ_0(t)v}\re^{iQ_0a}\rangle.
\eeqa
Here the brackets mean an average with respect  to $\rho_{th}$, and operators with a time argument are 
in the Heisenberg picture.  A somewhat simpler average is done fairly heroically in ref.\cite{NGK} using 
properties of Laguerre polynomials.   However, since the  sub-system and environment are all harmonic, a low-brow method is available.  A single simple harmonic oscillator obeys the well known Debye-Waller identity for thermal averages,
 \be
\big\langle{\re^{iqc}}\big\rangle = \re^{-\half c^2 \langle q^2\rangle },
\ee
where $q$ is the position operator and $c$ a number.  This is reviewed in the Appendix, where it is also shown that  a straightforward generalization yields \cite{err}
\beqa
\ln\mathcal T = & - & \half[(a+b)^2 + v^2]\langle Q_0^2\rangle -\half u^2\langle P_0^2\rangle 
-bu \langle Q_0 P_0(t)\rangle \nonu \\&-&bv\langle Q_0 Q_0(t)\rangle -uv\langle P_0 Q_0\rangle -ua\langle P_0(t) Q_0\rangle \nonu \\& -&av\langle Q_0(t) Q_0\rangle.
\eeqa
Since the co-ordinate $Q_0$ and the momentum $P_0$ are linearly related to the normal mode $q_\nu $s and $p_\nu$s via the known $X_{0\nu}$s the correlators in Eq.(18) are readily calculable, thereby formally completing the task of this section.

\section{IV. An example}

To illustrate the usefulness of these methods, consider $\langle Q_0^2\rangle$, one of the averages occurring in Eq.(18).  From Eq.(7)
\beqa
\langle Q_0^2\rangle &= &\sum_\nu \sum_{\nu^\prime} X_{0\nu} X_{0\nu^\prime} \langle q_\nu q_{\nu^\prime} \rangle = \sum_\nu X_{0\nu}^2 \langle q_\nu^2 \rangle \nonu \\ &=& \sum_\nu X_{0\nu}^2 \big({1 \over 2 \omega_\nu }\coth \half \beta \omega _\nu \big),
\eeqa
because the $\nu$s are independent oscillators, and $\langle q_\nu\rangle = 0$.   The sum can now be transformed using Eqs.(11) and (12).  It is easy to do analytically at zero  temperature ($\beta=\infty$).  Define the real part of the damped oscillator frequency via $\Omega^2=
\Omega_0^2 - \eta^2/4$ and factorize the denominator in Eq. (12) to obtain
\begin{eqnarray}
& &~~~~~~~~~~~~~~~~~~\langle Q^2_0\rangle_{T=0}\nonumber \\ &=&\frac{\eta}{2\Omega}\int_0^\infty \frac{d
\omega}{2 \pi}\bigg[ \frac{1}{(\omega -\Omega)^2 + \eta^2/4} -
\frac{1}{(\omega + \Omega)^2 +\eta^2/4}\bigg]\nonumber \\ &=& \frac{1}{2\Omega}\Big[ 1-
\frac{2}{\pi} \arctan \frac{\eta}{2\Omega} \Big].
\end{eqnarray}
The last is a known answer for the dissipation-induced squeezing by an ohmic bath. \cite{LC,va}
This effect is at the root of the effect of  damping in reducing the rate of escape from a metastable 
well.\cite{MAR}

\section{V. Time dependence of the probability}
The position-space probability for the sub-system $P(x,t)$ is obtained by setting $Q_{0f}^\prime =Q_{0f}^{\prime\prime} =x$ in Eq.(16), whereupon Eq.(15) requires that the variable $u \Rightarrow 0$ in Eq.(18).  The time dependence can then be completely described by 
\be 
\lan [Q_0(t) - Q_0(0)] Q_0(0)\ran\equiv -C(t) +i A(t),
\ee
this being the notation used in ref.\cite{VHVA} in the free-particle limit, with the sign as corrected in footnote\cite{err}.  One can now do the (gaussian) integrals in Eq.(16) to obtain, with $x^\prime \equiv Q_{0i}^\prime -Q_{0i}^{\prime\prime}, X^\prime \equiv \half (Q_{0i}^\prime + Q_{0i}^{\prime\prime})$,
\beqa
P(x,t)&= &\int dQ_{0i}^\prime dQ_{0i}^{\prime\prime} \alpha (Q_{0i}^\prime) \alpha (Q_{0i}^{\prime\prime}) J ( x , x^\prime, X^\prime, t),\nonu\\
J&=& {1\over 4\pi A(t) } {1\over \sqrt{2 \pi\lan Q_0^2 \ran}}\exp\big[ i {x^\prime\over 2 A(t)}(x - X^\prime )]\nonu\\ \times&\exp&  -\big [{{ x ^\prime}^2 C(t)\over 4 A^2(t)}+{1\over 2 \lan Q_0^2\ran}  (X^\prime -i{x^\prime  C(t)\over 2  A(t)})^2  \big].
\eeqa
In the free-particle limit $\Omega_0\rightarrow 0$, $\lan Q_0^2\ran \rightarrow \infty$ and the normalized gaussian in $X^\prime$ must be replaced, to preserve dimensions and normalization, by $L^{-1}$, where $L$ is the  size of the system, yielding
  \be 
J= {1\over 4\pi A(t) L}\exp{\big[ i {x^\prime\over 2 A(t)}(x - X^\prime )-{{ x ^\prime}^2 C(t)\over 4 A^2(t)}\big]}.
\ee
Using the method of Section II, known results for a free particle follow from Eq. (21):
\beqa
A(t) &=& {\eta\over \pi} \int_0^{\omega_c }d\omega  {\sin \omega t \over \omega (\omega ^2 +\eta ^2)}={1\over 2 \eta} (1 - \re^{- \eta t}),\nonu\\
C(t)&=&{\eta\over \pi}\int_0^{\omega_c} d\omega \coth {\beta \omega \over 2}{(1-\cos \omega t )\over \omega (\omega^2 + \eta^2)}\\ &\Rightarrow&(\beta\eta\gg 1)-{1\over \beta \eta^2}(1 - \re^{-\eta t} - \eta t)\nonu.
\eeqa
 \begin{figure}
   \includegraphics[width=\columnwidth]{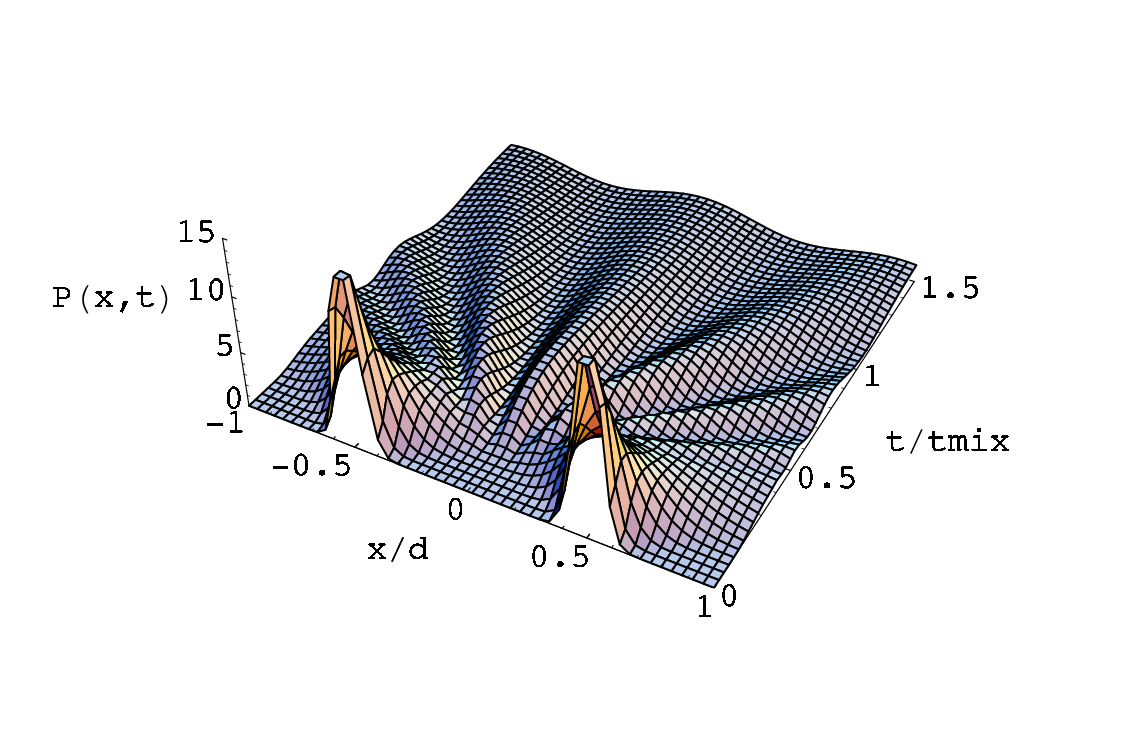}
  \caption{Time evolution at low temperature, $l_{th}/d =1$.  [The symbols are defined in the text.]}\label{fig1}
 \end{figure}
\begin{figure}
  \includegraphics[width=\columnwidth]{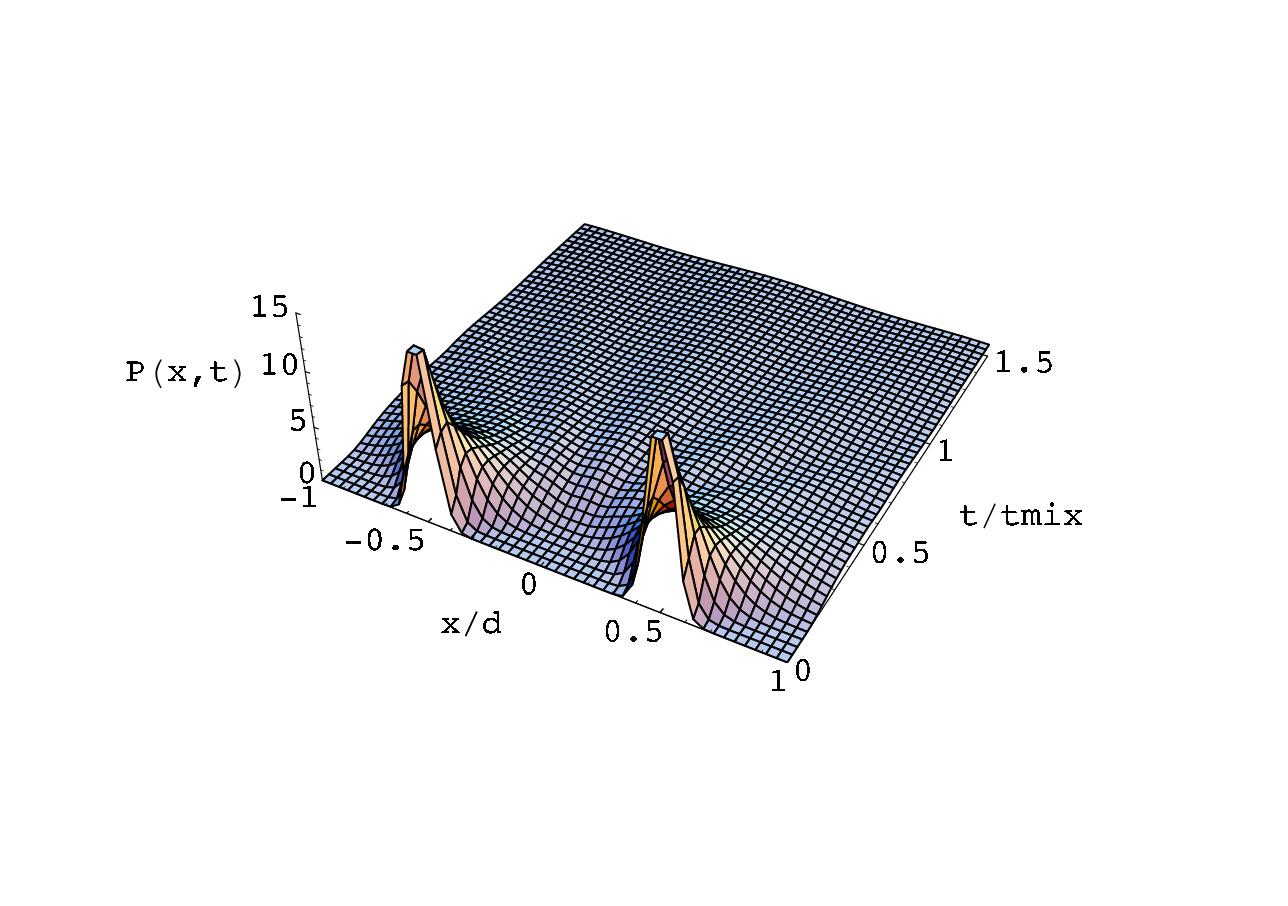}
\caption{Time evolution at moderate temperature, $l_{th}/d = 1/\sqrt 5$}\label{fig2}
 \end{figure}

  \section{VI  Decoherence}
  
 Eq.(22)---which is not restricted to ohmic dissipation---is to my knowledge new, and explicit enough to allow a general study of time development in this dissipative quantum system.  Previous attempts \cite{VA,GOA} in which I have been involved have for technical reasons been restricted to high temperature, a limitation which it should here be  possible to avoid.   Work, in progress, in this direction would seem to be justified by experimental\cite{MOOIJ} and theoretical\cite{Grifoni}  interest in quantum information storage.
 
 The free particle limit, Eqs.(23,24), has been derived by many different methods---perhaps none as straightforward as the one here given.  At finite $\eta$, these equations (uncontroversially) display decoherence. This is particularly well demonstrated in an example introduced in ref\cite{FLO}.  Here the aperture function is taken to be a sum of two Gaussians, each of width $\sigma$ and separated by a distance $d$.  The probability given by Eqs. (23) and (24)
can then be written as a sum of the probabilities from each slit alone (sum term) and an interference contribution depending sinusoidally on a time dependent phase (interference term.)   
There is an unresolved controversy \cite{GvDA, FOC} in the published literature about what is meant by decoherence in this  completely well defined problem.    
  No  one would doubt that the amplitude of the interference term is a measure of coherence.   In
ref.\cite{FLO} and other publications\cite{FC}  an ``attenuation coefficient" is introduced which is equivalent to the amplitude of the interference term divided by the sum term, evaluated at the mid-point  between the slits.  This quantity is by construction unity at $t=0$.  It decreases rapidly with time.   At high temperatures, in a  sense to be made clear below, it drops to zero, even when the environmental coupling is eliminated.  This is interpreted in refs.\cite{FLO, FC} as ``Decoherence without Dissipation."  
\begin{figure}
 \includegraphics[width=\columnwidth]{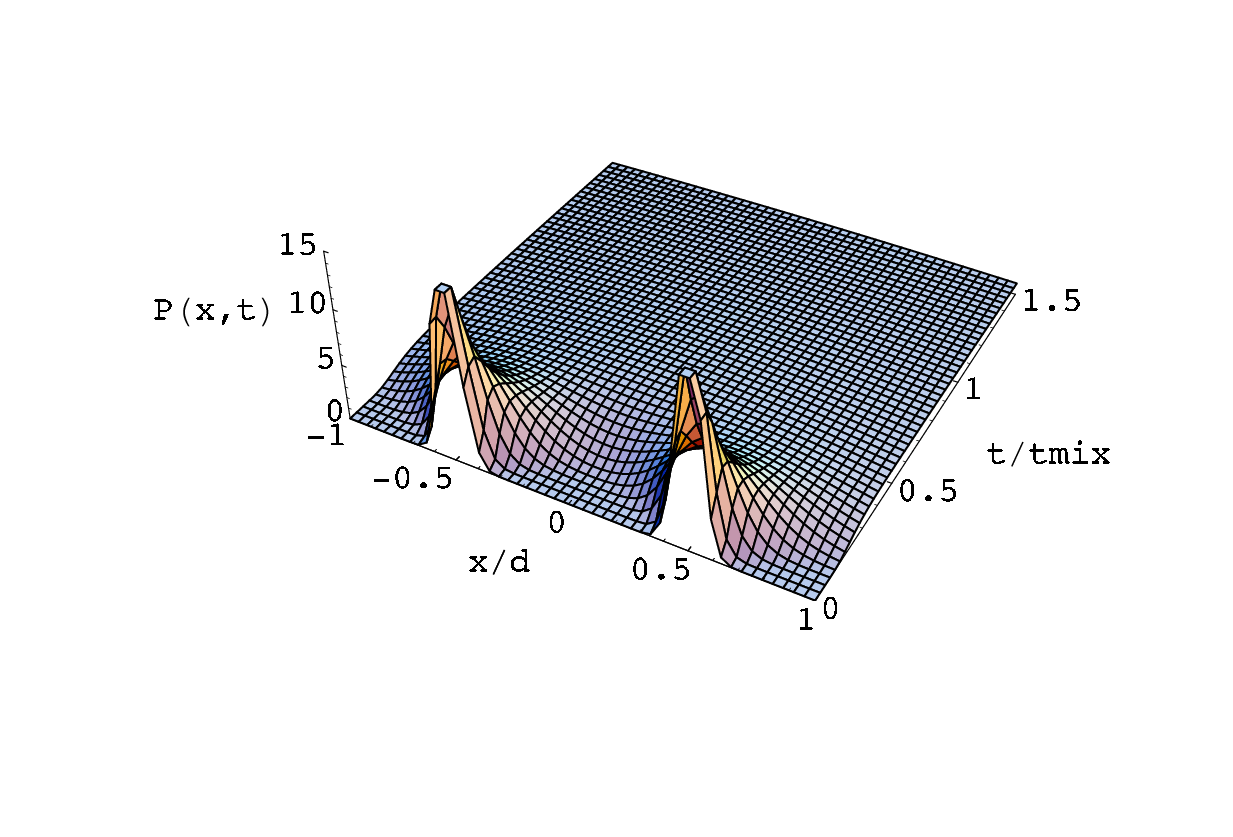}
  \caption{Time evolution  at high temperature, $l_{th}/d = 1/5$}\label{fig3}
 \end{figure}

In a Comment \cite{GvDA}  it is suggested that the measure of decoherence used in
 \cite{FLO} does not distinguish between the loss of phase information and the spreading of wave packets on time scales less than the mixing time $t_{mix} = 2 \sigma d (m/\hbar )$.  Here $\sigma$
is the width of each slit, $d$ the spacing between them, and I have reintroduced Planck's constant $\hbar$ and the particle mass $m$ to make the dimensions transparent.

 This has been vigorously rebutted in a Reply\cite{FOC}.  
  
 Rather than repeat these arguments, I refer the interested reader to them.  However, since many readers may be intrigued by the idea of decoherence without dissipation, I  close this paper with 4 figures which show that there is no evidence for any such thing in the uncontroversial Eqs.(23,24).    Since the disagreement occurs in the limit of no dissipation, consider this case at various temperatures---given by the  ratio of the thermal de Broglie wavelength $l_{th}= \sqrt{\beta/ 2} (\hbar/\sqrt m)$ to $d$.  The figures, in which $\sigma/d = 0.05$, show that at low temperatures,  $l_{th}\sim d$ there is coherence without decoherence.  At higher temperatures, there is no coherence at all, even on the short time scale
\be
\tau_{FLO}=\sqrt {8\beta m} {\sigma^2\over d}= 2 t_{mix} \cdot{l_{th}\over d}\cdot{\sigma\over d}
\ee
 introduced in \cite{FOC},  and thus nothing to  decohere.  In this limit, coherence is already destroyed by the Hakim-Ambegaokar initial condition.
\begin{figure}
 \includegraphics[width=\columnwidth]{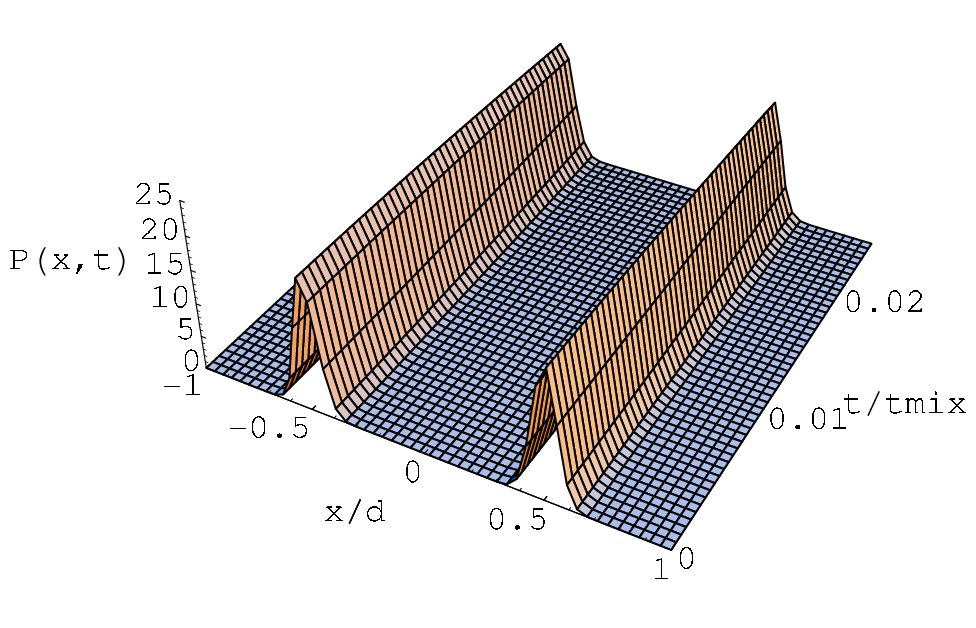}
  \caption{Blow-up of Fig.3 on the scale of $\tau_{FLO}=0.02~ t_{mix},$ for the chosen parameters.  }\label{fig4}
 \end{figure}

\noindent \underbar {Added Note.}  After this paper was completed, I was made aware of \cite{ FMDG} in which normal co-ordinates are used to treat
 the problem of many non-interacting fermions  coupled to a disspative environment in a harmonic oscillator.
 \section{Appendix}
 A wonderfully short  proof of the identity for thermal averages
 \be
 \lan \exp [ \sum_i  d_i  a_i +  c_i  a^\dagger_i ] \ran = \exp [ \half \sum_i  \lan (d_i a_i + c_i a^\dagger_i)^2\ran ],
 \ee
 where $a_i $, $a^\dagger_i$ are boson annihilation and creation operators and the subscript $i$
 refers to independent harmonic oscillators, is given in a single mannerist, if not rococo, sentence by Mermin\cite{NDM}.  Note that, since all the operators in Eq.(16) have c-number commutators, it  can be put in the form of the left hand side of Eq.(26) for the subsystem oscillator labeled $0$, using the Baker-Haussdorf identity for such operators: 
 $\re^A \re ^B = \re^{[A +B]}\re^{{1\over 2} [A, B]}$. Now, express $P_0$ and $Q_0$ in terms of the (normal) co-ordinates, of the independent oscillators $\nu$.   The Debye-Waller identity
 Eq. (26) then has in the exponent a sum over correlators for each $\nu$.   Note that as in Eq. (19), a single sum can be replaced by a double sum, because $\lan q_\nu\ran=  \lan  p_\nu\ran =0$, yielding Eq.(18).

 \section {acknowledgements}

The figures in this paper were created using Mathematica by Dominique Gobert.  I am grateful to him and Jan von Delft for considerable help, and to Frank Wilhelm for background information.  This work was started during the summer of 2004 at the Aspen Center for Physics. where I had useful discussions with Vladimir Privman and Dima Mozyrsky.  Support from the NSF under grant DMR- 0242120 is acknowledged with thanks.


\begin{thebibliography}{99}
  
 \bibitem{VHVA} 
  V. Hakim and V. Ambegaokar, Phys. Rev. A {\bf 32}. 423 (1985).
  
  \bibitem{FOC}
  G. W. Ford and R.F. O'Connell,  Phys. Rev. A  {\bf 70}, 026102 (2004).
  
  \bibitem{GvDA}
  D. Gobert, J. von Delft, and V. Ambegaokar, Phys. Rev. A {\bf 70}, 026101 (2004).  See also arXiv:quant-ph/0306019 by the same authors.
   
   \bibitem{NGK}
   N. G. van Kampen, J. Stat. Phys. {\bf 115}, 1057 (2004).   Similar methods are used in G. W. Ford, J. T. Lewis, and R. F O'Connell, J. Stat. Phys. {\bf 53}, 439 (1988).  Citations of the considerable early literature on heat baths treated as continua of harmonic oscillators are contained in these two references.
   \bibitem{VA}
   V. Ambegaokar, Ber.  Bunsenges. Phys. Chem {\bf 95} 400 (1991).
      \bibitem{SP}
   H.Spohn, Rev. Mod. Phys {\bf 52} 569 (1980).

    \bibitem {LC}
    A. O. Caldeira and A. J. Leggett, Physica {\bf 121A}, 587 (1983).
    \bibitem{va}
    See, e.g., V. Ambegaokar, in G. T. Moore and M. O. Scully eds., NATO ASI
Series B: Physics Vol.135, 231 (1984).
\bibitem{err}
I have verified that in the free-particle limit this equation reproduces Eqs.(35a,b) of ref.\cite{VHVA} except for an error of sign in Eq.(34a) and Appendix C.  With this change of sign, $A(t)$ is positive and given by Eq.(24) in the present paper, Eq.(38)  is correct,  but the second form of Eq.(39) now has a positive sign.  This error was noted in \cite{FLO}
 
\bibitem{FLO}  G. W. Ford, J. T. Lewis, and R. F O'Connell, Phys. Rev A {\bf 64}, 032101 (2001).
 
 \bibitem{MAR}
 J. Martinis, M.H. Devoret, and J. Clarke, Phys.Rev. B {\bf 35}, 4682 (1987).  
 \bibitem{GOA}
 A. Garg, J. Onuchic, and V. Ambegaokar J. Chem Phys. {\bf 83} 4491 (1985).
 \bibitem{MOOIJ}
 I. Chiorescu et al. Nature {\bf 431} (7005), 159 (2004).
 \bibitem{Grifoni}
 M. Thorwant et al. Chem. Phys. {\bf 296},  333 (2004).
 \bibitem{FC}
 G.W. Ford and R.F. O'Connell,  Phys. Lett. {\bf 286}, 87 (2001); Am. J. Phys. {\bf 70}, 319 (2002). 
 \bibitem{NDM}
  N.D. Mermin, J. Math. Phys. {\bf 7}, 1038 (1966).


\bibitem{FMDG}
F. Marquardt and D.S. Golubev, ArXiv:cond-mat/0409401.



   \end{thebibliography}
\enddocument